\documentclass{PoS}
\usepackage{psfrag}
\usepackage{amssymb}
\usepackage{epsfig}
\usepackage{amsmath}
\usepackage{amsfonts}
\usepackage{dcolumn}  

\title{The First Moment of the Kaon Distribution Amplitude from $N_f=2+1$ Domain Wall Fermions\footnotetext{SHEP-0629}}

\ShortTitle{The First Moment of the Kaon Distribution Amplitude from $N_f=2+1$ Domain Wall Fermions}

\newcommand{\ovra}[1]{\overset{\rightarrow}{#1}}
\newcommand{\ovla}[1]{\overset{\leftarrow}{#1}}
\newcommand{\ovlra}[1]{\overset{\leftrightarrow}{#1}}

\newcommand{\xiav}{\langle\,\xi\,\rangle}
\parskip\smallskipamount
\def\firstmomentop{O_{\{\rho\mu\}}}
\def\MSbar{{\ensuremath{\overline{\mathrm{MS}}}}}
\def\latt{\mathrm{latt}}
\def\bigG{\frac{g^2 C_\mathrm{F}}{16\pi^2}}
\def\Mmf{M^{\mathrm{MF}}}
\def\wmf{w_0^{\mathrm{MF}}}
\def\Tmf{T_\mathrm{MF}}
\def\gev{\,\mathrm{Ge\kern-0.1em V}}

\author{
	P.A.~Boyle$\,^a$, M.A.~Donnellan$\,^b$, J.M.~Flynn$\,^b$, 
	\speaker{A.~J\"uttner}$\,^b$, J.~Noaki$\,^b$, C.T.~Sachrajda$\,^b$, 
	R.J.~Tweedie$\,^a$\\
        $^a$ Department of Physics and Astronomy, University of
	Edinburgh, Edinburgh, EH9 3JZ, UK\\
        $^b$ School of Physics and Astronomy, University of Southampton,
	Southampton, SO17 1BJ, UK\\
        E-mail: \email{juettner@phys.soton.ac.uk}\\
	{\bf UKQCD Collaboration}
	}

\abstract{We present a lattice computation of the first moment of the kaon's
leading-twist distribution amplitude. We use 
ensembles with 2+1 dynamical flavours of domain wall
fermions and the Iwasaki gauge action from the RBC and UKQCD
joint dataset. We observe the expected chiral
behaviour and obtain $\xiav(2\,\textrm{GeV})\equiv 3/5
\,a_K^1\,(2\,{\rm GeV})=0.032(3)$, which agrees very well with 
other results obtained using QCD sum-rules and the recent lattice
result from the UKQCD/QCDSF collaboration.}

\FullConference{XXIVth International Symposium on Lattice Field Theory\\
                July 23-28, 2006\\
                Tucson, Arizona, USA}

\begin{document}

\section{Introduction}
We present a lattice calculation of the first moment of the
leading-twist distribution amplitude of the kaon, $\phi_K(u,\mu)$
\cite{Boyle:2006pw}.
Among the many phenomenological applications which require knowledge
of distribution amplitudes are  electromagnetic
form-factors at large momentum transfer and related processes
\cite{Chernyak:1977as,Chernyak:1980dj,
Efremov:1979qk,Efremov:1978rn,Chernyak:1977fk,Chernyak:1980dk,Lepage:1980fj},
and, following the development of the factorization
framework, exclusive charmless two-body
$B$-decays  into two light mesons
~\cite{Beneke:1999br,Beneke:2000ry,Beneke:2001ev,
Bauer:2000ew,Bauer:2000yr,Bauer:2001ct,Bauer:2001yt}.\\
The distribution amplitude
pa\-ra\-me\-trizes the overlap of a kaon with longitudinal
momentum $p$ with the lowest Fock state consisting of a quark and
an anti-quark carrying the momentum fractions $up$ and $\bar
up=(1-u)p$, respectively ($u+\bar u=1$). It is defined by the
non-local (light-cone) matrix element
\begin{equation}\label{eq:phidefz}
 \left.\langle\,0\,|\,\bar{q}(z)\,\gamma_\rho\gamma_5\,{\cal
 P}(z,-z)\,s(-z)\,|\,K(p)\,\rangle
 \right|_{z^2=0}\equiv{f_K}\,(ip_\rho)\,
 \int_0^1du\,e^{i(u-\bar u)p\cdot z}\phi_K(u,\mu)\,,
\end{equation}
where $\mu$ is a renormalization scale and
${\cal P}(z,-z)={\cal
 P}\,\exp\left\{-ig\int_{-z}^{z}dw^\mu A_\mu(w)\right\}$.
The distribution
amplitude is normalized by $\int_0^1du\,\phi_K(u,\mu)=1$ and can
be expanded in terms of Gegenbauer polynomials $C^{3/2}_n(2u-1)$,
\begin{equation}
 \phi_K(u,\mu)=6u\bar u\left(1+\sum\limits_{n\ge 1}a_n^K(\mu)
    C_n^{3/2}(2u-1)\right).
\end{equation}
The lowest Gegenbauer moment $a_1^K$
is proportional to the average difference of the
longitudinal quark and anti-quark momenta of the lowest Fock
state,
\begin{equation}\label{eq:define_a1K}
a_1^K(\mu)={5\over 3}\int_0^1 du (2u-1)\,\phi_K(u,\mu)
=\frac53\,\langle 2u-1\rangle\equiv\frac53\,\xiav(\mu)\,.
\end{equation}
While the first moment of the distribution amplitude vanishes in the case
of the pion, it  
is non-zero for the Kaon because of SU(3)-breaking effects.
$\xiav$ is obtained from the matrix element of a local
operator,
\begin{equation}\label{eq:1st_m_def_cont}
 \langle\,0\,|\,\bar{q}(0)\,\gamma_\rho\gamma_5
 \,\ovlra{D}_\mu
 \,s(0)\,|\,K(p)\,\rangle = \xiav\,f_K\,p_\rho \,p_\mu=
 {3 \over 5}\,a_1^K f_K\,p_\rho \,p_\mu\,,\\
\end{equation}
where we use
$\ovlra{D}_\mu=\ovla{D}_\mu-\ovra{D}_\mu$, $\overset{\rightarrow}
{D}_\mu=\overset{\rightarrow}{\partial}_\mu+ig A_\mu$ and
$\overset{\leftarrow}
{D}_\mu=\overset{\leftarrow}{\partial}_\mu-ig A_\mu$.\\
The first moment of the kaon's distribution amplitude has in the
past been determined mainly from QCD sum rules, and recent results
include:
$a_1^K(1\,{\rm GeV}) \,=\,0.05(2)$ \textrm{\cite{Khodjamirian:2004ga},}
$0.10(12)$ \textrm{\cite{Braun:2004vf}},
$0.050(25)$ \textrm{\cite{Ball:2005vx}} and
$0.06(3) \textrm{\cite{Ball:2006fz}}$.
Very recently an independent lattice study of this quantity was published~
\cite{Braun:2006dg}
which quotes $a_1^K(2{\rm GeV})=0.0453\pm0.0009\pm0.0029$ as the final result.\\
Here we use the $N_f=2+1$ gauge field
ensembles from the RBC and UKQCD dataset \cite{BobProc,RobProc,configpaper} 
(domain wall fermions 
\cite{Kaplan:1992bt,Furman:1994ky} and Iwasaki gauge action 
\cite{Iwasaki:1984cj,Iwasaki:1985we})
with three values of the light-quark  mass 
with $m_{\rm sea}=m_{\rm valence}$ in each case.
The hadronic spectrum and
other properties of these configurations have been presented at this conference
\,\cite{BobProc,RobProc,configpaper}.

\section{\boldmath{$\xiav^{\bf bare}$ from Lattice Correlation Functions}}
\label{sec:Lattice study}
In constructing the lattice operators which are 
relevant for the determination of
$\xiav$, we use the following
symmetric left- and right-acting covariant derivatives:
\begin{eqnarray}
 \overset{\rightarrow}{D}_\mu\psi(x)=\frac{1}{2a}\left\{\,U(x,x+\hat\mu)
    \psi(x+\hat\mu)- U(x,x-\hat\mu)\psi(x-\hat\mu)\,\right\}\,,\\
 \bar\psi(x)\overset{\leftarrow}{D}_\mu=\frac{1}{2a}\left\{\bar{
 \psi}(x+\hat\mu)U(x+\hat\mu,x)- \bar{
 \psi}(x-\hat\mu)U(x-\hat\mu,x)\,\right\}\,,
\end{eqnarray}
where the $U$'s are the gauge links and $\hat\mu$ is a vector of
length $a$ in the direction $\mu$ ($a$ denotes the lattice
spacing).\\
To illustrate the method, consider the local lattice
operators 
$O_{\rho\mu}(x)=\bar{q}(x)\gamma_\rho\gamma_5\,\ovlra{D}_\mu s(x)$, 
$A_{\rho}(x)=\bar{q}(x)\,\gamma_\rho\gamma_5 s(x)$ and
$P(x)=\bar{q}(x)\,\gamma_5\, s(x)$
from which we define the two-point correlation functions
\begin{equation}\label{eq:define_C_rhomu}
 \begin{array}{rclrcl}
  C_{\rho\mu}(t,{\vec p}\,)=\sum\limits_{\vec x}e^{i \vec{p}\cdot\vec{x}}
  \langle 0|O_{\rho\mu}(t,\vec{x}\,) P^\dagger(0)|0\rangle&{\rm and}&
  C_{A_\nu P}(t,\vec{p}\,)=\sum\limits_{\vec x}e^{i\vec{p}\cdot\vec{x}}
  \langle 0|A_\nu(t,\vec{x}\,) P^\dagger(0)|0\rangle\,.
 \end{array}
\end{equation}
Here $q$ and $s$ represent the light and strange quark fields,
respectively.
At large Euclidean 
times $t$ and $T-t$ ($T$ is the length of the lattice in the 
time direction), we expect
\begin{equation}\label{eq:ratio_R}
 R_{\{\rho\mu\};\,\nu}(t,\vec{p}\,)\,\equiv\,\frac
 {C_{\{\rho\mu\}}(t,\vec{p}\,)}
 {C_{A_\nu P}(t,\vec{p}\,)}
\,\to\,
 \,i\,{p_\rho p_\mu \over p_\nu}\,\xiav^\textrm{bare}\,.
\end{equation}
The superscript \textit{bare} denotes the fact that the operators are
the bare ones in the lattice theory with ultraviolet cut-off
$a^{-1}$ in the Domain Wall Formalism
and the braces in the subscripts $\{\rho\mu\}$ indicate that the 
indices are symmetrized.
In order to avoid mixing of $O_{\mu\nu}$ under 
renormalization \cite{Gockeler:1996mu}
we only consider the cases $\rho=\nu=4$, $\mu=k$ ($k=1,2,3$) 
with
$p_k=\pm2\pi/L$ while $|\vec p|=2\pi/L$.

\section{Perturbative Renormalization of the Lattice Operators}
\label{sec:Renormalization}

The perturbative matching from the lattice to the $\MSbar$ scheme is
performed by comparing one-loop calculations of the 
two-point Green function with an insertion of the operator
$\firstmomentop$ in both schemes. 
Defining $\firstmomentop^\MSbar(\mu) = Z_{\firstmomentop}
\firstmomentop^\latt(a)$, the renormalization factor is given by
\begin{equation}
\label{eq:Zpt} Z_{\firstmomentop} = \frac1{(1-w_0^2)Z_w}
 \left[ 1 + \bigG \left( -\frac83 \ln(\mu^2 a^2) + \Sigma_1^\MSbar
        - \Sigma_1 + V^\MSbar - V \right) \right].
\end{equation}
In this expression, $(1-w_0^2)Z_w$ is a characteristic
normalization factor for the physical quark fields in the domain
wall formalism. It is a common factor in the numerator and
denominator of the ratio $R_{\{\rho\mu\};\nu}$ as are the
contributions from the wave function renormalization. $Z_w$
represents an additive renormalization of the large Dirac mass or
domain wall height $M=1-w_0$ which can be rewritten in
multiplicative form at one-loop as
$Z_w = 1+\bigG z_w$ with  $z_w = \frac{2w_0}{1-w_0^2}\,\Sigma_w$.

The terms $\Sigma_1^\MSbar$ and $\Sigma_1$ come from quark wave
function renormalization. The terms $V^\MSbar$ and $V$ come from
the one-loop corrections to the amputated two-point function. 
Using naive dimensional
regularisation in Feynman gauge with a gluon mass infrared
(IR) regulator,
$\Sigma_1^\MSbar = \frac12$ and $V^\MSbar = -\frac{25}{18}$.\\
The contribution $\Sigma_1$ has been evaluated for domain
wall fermions with the Iwasaki gluon action in Feynman gauge in~\cite{Aoki:2002iq}. We have calculated
the lattice vertex term $V$ for the same action and gauge 
regulator to complete the evaluation of $Z_{\firstmomentop}$. The
perturbative calculation is explained
in~\cite{Aoki:1998vv,Aoki:2002iq,Capitani:2005vb} and the form of
the Iwasaki gluon propagator can be found
in~\cite{Iwasaki:1983ck}.\\
For the Iwasaki
gluon action and for the value of $M=1.8$ used here
the physical quark normalization $z_w$
has been found to be 
very large in \cite{Aoki:1998vv,Aoki:2002iq} and we therefore use mean
field improvement as described in \cite{Aoki:2002iq}.\\
The first step is to define a mean-field value for the domain wall
height,
$\Mmf = M - 4(1-P^{1/4})$
where $P=0.58813(4)$ is the average plaquette in our
simulations, leading to
$\Mmf = 1.3029$.
The physical quark normalization factor becomes
$\left[1-(\wmf)^2\right]Z_w^\mathrm{MF}$, with
$Z_w^\mathrm{MF} = 1+\bigG z_w^\mathrm{MF}$ and 
$z_w^\mathrm{MF} = \frac{2\wmf}{1-(\wmf)^2}\,(\Sigma_w + 32\pi^2\Tmf) = 5.2509$,
where $\Tmf=0.0525664$~\cite{Aoki:2002iq} is a mean-field tadpole
factor and $\Sigma_w$ is evaluated at $\Mmf$.
Likewise, $\Sigma_1=3.9731$ and
$V=-4.1907$ in equation~(\ref{eq:Zpt}) are evaluated at $\Mmf$ and
the mean-field improved renormalization factor for our simulations
becomes:
\begin{equation}
Z_{\firstmomentop} = \frac1{0.9082}\,
 \left[1-\bigG \, 5.2509\right]
 \left[1+\bigG \left( -\frac83 \ln(\mu^2 a^2) -0.6713\right) \right].
\end{equation}
We make two choices for the mean-field improved $\MSbar$ coupling.
The first uses the measured plaquette value, $P$, according
to~\cite{Aoki:2002iq}
\begin{equation}
\frac1{g^2_\MSbar(\mu)} =
 \frac P{g^2} + d_g + c_p + \frac{22}{16\pi^2}\,\ln(\mu a)\,,
\end{equation}
where $d_g=0.1053$ and $c_p=0.1401$ for the Iwasaki gauge action
and $\beta = 6/g^2 = 2.13$ in our simulations. The second choice
is the usual continuum $\MSbar$ coupling. At $\mu a = 1$, we find
$\alpha_\MSbar(\mathrm{plaq}) = 0.1752$ and
$\alpha_\MSbar(\mathrm{ctm}) = 0.3385$. 
With these two choices of coupling, our
value for the renormalization factor becomes
\begin{equation}\label{eq:zrpert}
\frac{Z_{\firstmomentop}}{Z_\mathrm{A}} =  \begin{cases}
 1.2346 & \mbox{plaquette coupling}\\
 1.3384 & \mbox{continuum \MSbar}\ .
             \end{cases}
\end{equation}
We include the spread of results
in eq.(\ref{eq:zrpert}) as the estimate of our current systematic
uncertainty in the renormalization factor 
and thus we will eventually use 
$\frac{Z_{\firstmomentop}}{Z_\mathrm{A}}=1.28\pm 0.05$
for the final result.

\section{Numerical Simulation and Results}\label{sec:Results}

The lattice volume is $(L/a)^3\times T/a\times L_s=16^3\times32\times 16$.
The choice of bare
parameters is $\beta=2.13$ for the gauge coupling,
$am_s=0.04$ for the strange quark mass (which has been tuned to
correspond to the physical value) and $am_q=0.03,\,0.02,\,0.01$
for the light-quark masses. With these simulation
parameters the lattice spacing is
$a^{-1}=1.60(3)$\,GeV~\cite{RobProc,configpaper}. Owing to the remnant chiral
symmetry breaking the quark mass has to be corrected additively by
the residual mass in the chiral limit, $am_{\rm
res}=0.00308(3)$~\cite{RobProc,configpaper}.\\
%
\subsection{Bare correlation functions}
For each value of the light-quark mass we computed the correlation
functions on 300 gauge configurations separated by 10 trajectories
in the Monte Carlo history. On each configuration we average the
results obtained from 4 positions of the source for the lightest quark mass
($am_q=0.01$) and 2 positions of the source for the remaining two masses
($am_q=0.02$ and 0.03). 
In order to improve
the overlap with the ground state at the source where we insert
the density $P^\dagger$, we employed gauge invariant Jacobi
smearing \cite{Allton:1993wc} (radius 4 and 40 iterations) with
APE-smeared links in the covariant Laplacian operator ($4$ steps
and smearing factor $2$) \cite{Falcioni:1984ei,Albanese:1987ds}.

The kaon masses corresponding to the simulated bare
light-quark masses are $am_K^{0.03}=0.4164(10)$,
$am_K^{0.02}=0.3854(10)$, and
$am_K^{0.01}=0.3549(14)$.

\begin{figure}
 \begin{center}
 \begin{minipage}{.49\linewidth}
  \psfrag{tovera}[t][c][1][0]{$t/a$}
  \psfrag{1stmoment}[c][t][1][0]{$\xiav^{\textrm{bare}}$}
  \psfrag{Legendmass1}[c][c][1][0]{\scriptsize$am_{ud}=0.01$}
  \psfrag{Legendmass2}[c][c][1][0]{\scriptsize$am_{ud}=0.02$}
  \psfrag{Legendmass3}[c][c][1][0]{\scriptsize$am_{ud}=0.03$}
  \epsfig{scale=.27,angle=270,file=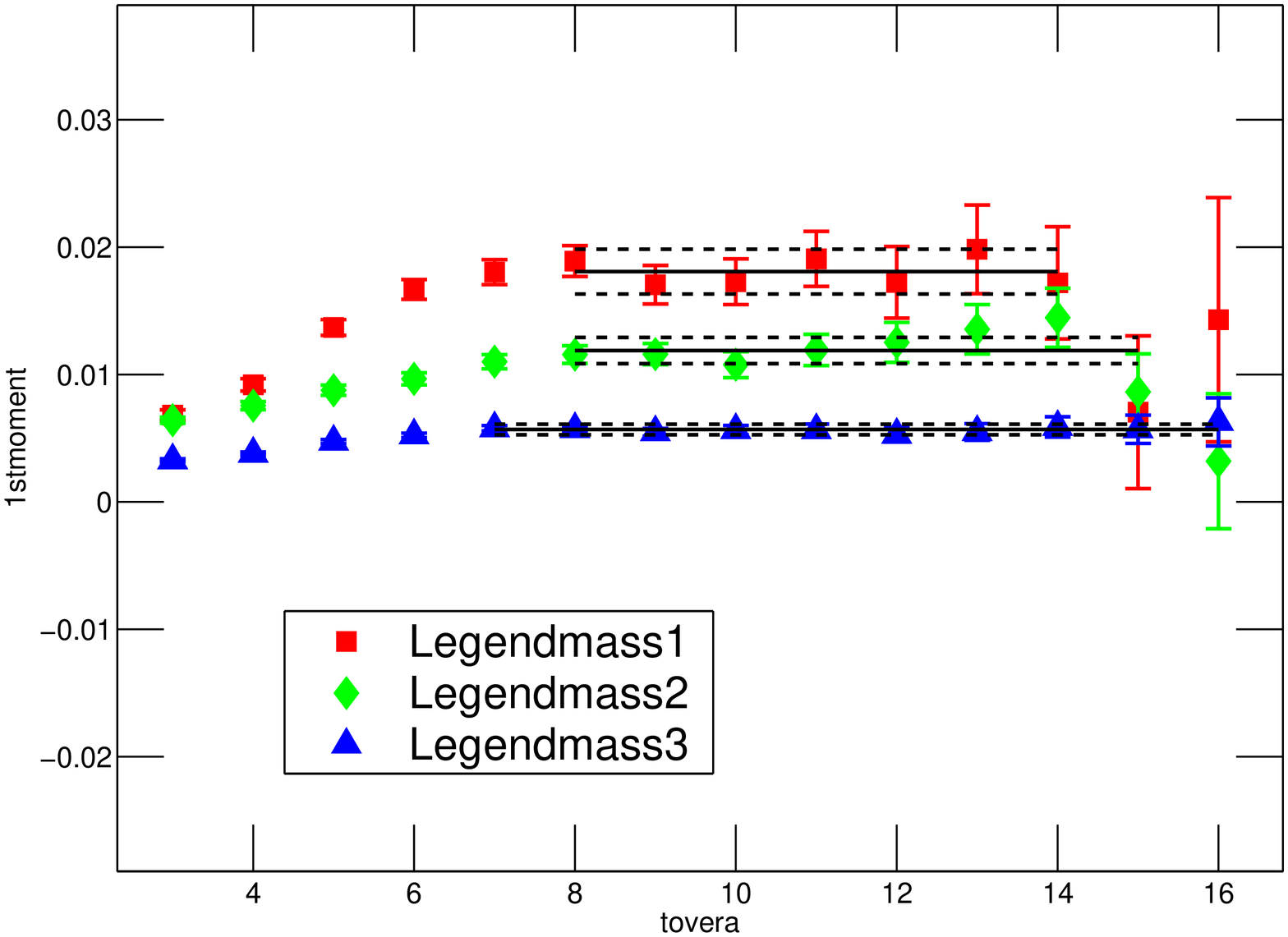}\\
 \end{minipage}
 \begin{minipage}{.49\linewidth}
  \psfrag{mqplusmres}[t][c][1][0]{$am_q+am_{\rm res}$}
  \psfrag{fstmom}[c][t][1][0]{$\xiav^{\textrm{bare}}$}
  \psfrag{Legendmass1}[c][c][1][0]{{\scriptsize$am_{ud}=0.01$}}
  \psfrag{Legendmass2}[c][c][1][0]{\scriptsize$am_{ud}=0.02$}
  \psfrag{Legendmass3}[c][c][1][0]{\scriptsize$am_{ud}=0.03$}
  \epsfig{scale=.27,angle=270,file=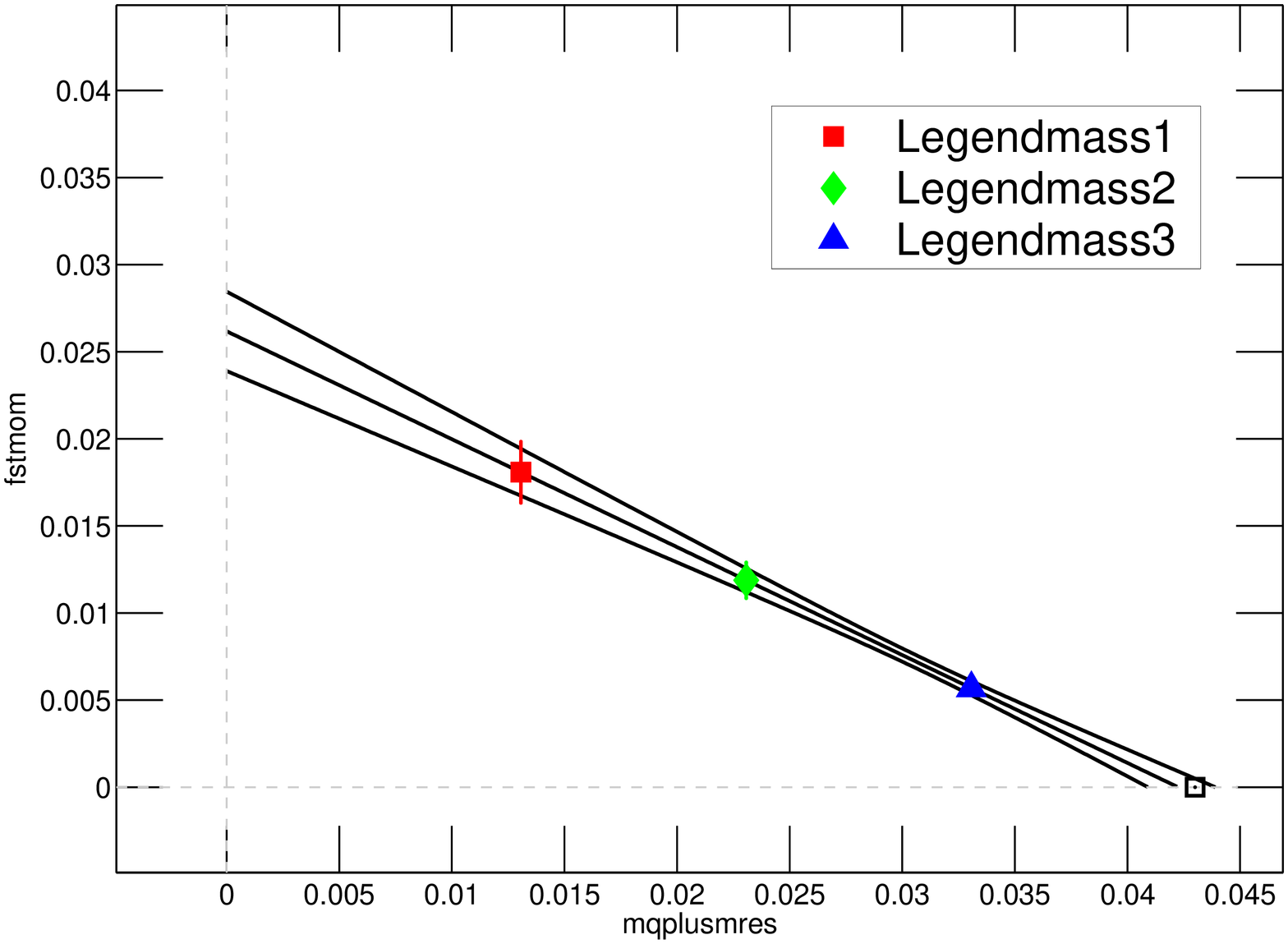}\\
 \label{fig:1st_ch_extrapol}
 \end{minipage}
 \end{center}
 \caption{Left: Jack-knife results for $\xiav^{\textrm{bare}}$ as a 
	function of the time. The ranges over which we fit and the 
	corresponding results are indicated by the black lines.
	Right: Linear chiral extrapolation for $\xiav^{\textrm{bare}}$.}

	\label{fig:1st_cf}
 \label{fig:1st_ch_extrapol}
\end{figure}

The left plot in figure \ref{fig:1st_cf} shows our results for $\xiav^{\rm bare}$
as a function of $t$ obtained from  the ratio $R_{\{4
k\};\,4}(t,p_k=\pm 2\pi/L)$ for the three values of the mass of
the light quark. 
We averaged the results over equivalent choices for the momenta and folded
the data in the time-direction.
There are clear plateaus, demonstrating that the
$SU(3)$-breaking effects are measurable and $\xiav$ can be
determined. 

\subsection{Chiral extrapolation}\label{subsec:chiral}

Plotting our results for $\xiav^\textrm{bare}$ as a function of the
light-quark mass in the right plot in fig.~\ref{fig:1st_ch_extrapol} 
and taking into
account the remnant chiral symmetry breaking by defining the
chiral limit at the point $am_q+am_{\rm res}=0$ our data confirms the linear
behaviour predicted by chiral perturbation 
theory~\cite{Chen:2003fp,Chen:2005js}.
Moreover the line passes through $\xiav^\textrm{bare}=0$ at a value of the
light-quark mass (denoted by the open square) 
which is consistent with the mass
of the strange quark, as expected for the $SU(3)$ symmetric case
($am_{ud}=am_{s}=0.04$). 
From the linear fit we obtain $\xiav^\textrm{bare}=0.0262(23)$ in
the chiral limit.

\section{Systematic Uncertainties and our Final
Result}\label{sec:final} 
Combining $\xiav^{\rm bare}$ with the result for the perturbative 
renormalization factor we obtain our final result
\begin{equation}\label{eq:resulta}
\xiav^{\MSbar}(\mu=1.6\,\textrm{GeV})=0.034\pm
0.003\,.\end{equation}

In order to compare our result with previous
calculations we evolve it to the renormalization scales 1\,GeV and
2\,GeV using the three-loop anomalous
dimension~\cite{Larin:1993vu}. We obtain
$\xiav^{\MSbar}(\mu=2\,\textrm{GeV})=0.032\pm 0.003$ and 
$\xiav^{\MSbar}(\mu=1\,\textrm{GeV})=0.040\pm 0.004$.\\
The error in the renormalization factor due to the uncertainty in
the lattice spacing is negligible. For example if we
conservatively allow the lattice spacing to vary between 1.58\,GeV
and 1.62\,GeV, the contribution to the relative error on
$\xiav^{\MSbar}$ is less than 0.2\%.

Among the uncertainties which we are not at this stage
in a position to check numerically are the continuum
extrapolation, finite-volume effects and the fact that the strange
quark mass ($m_sa=0.04$) is only approximately tuned to its
physical value. The lattice artefacts are formally of
$O(a^2\Lambda^2_{\textrm QCD})\simeq 2.5\%$ and we
are planning to check this with a simulation at a smaller lattice spacing.
 We would expect the
finite volume effects to be small and are currently checking this
with a simulation on a $24^3\times 64$ lattice. The strange quark
mass appears to be well tuned~\cite{RobProc,configpaper} so again we
expect the contribution to the error from this uncertainty to be
very small. Thus we expect the errors from these three sources to
be sufficiently small not to change the errors quoted for our final result.
We are also carrying out a systematic
programme of non-perturbative renormalization which will enable us
to reduce the uncertainty in the renormalization constants.

\section{Summary and Conclusions}\label{sec:concs}
We have demonstrated that the
$SU(3)$-breaking effects which lead to a non-zero value for the
first moment of the kaon's distribution amplitude are sufficiently
large to be calculable in lattice simulations and satisfy the
expected chiral behaviour. 
As our best result we quote $\xiav^{\MSbar}(\mu=1.6\,\textrm{GeV})=0.034\pm
0.003$.\\[2mm] 
{\bf Acknowledgements} \\

The development and computer equipment used in this calculation
were funded by the U.S. DOE grant DE-FG02-92ER40699, PPARC JIF
grant PPA/J/S/1998/00756 and by RIKEN. This work was supported by
PPARC grants PPA/G/O/2002/00465, PPA/G/S/\-2002/00467 and
PP/D000211/1. JN acknowledges support from the Japanese Society
for the Promotion of Science.

\bibliography{kaon_da}
\bibliographystyle{elsevier}

\end{document}